\documentclass[11pt]{article}
\usepackage{amsmath,amssymb,epsfig,pifont,verbatim,natbib,amsthm,bm}
\graphicspath{{./}{../graphs/}{../pit-graphs/}}
\newcommand{\ud}{\mathrm{d}}
\bibpunct{(}{)}{;}{a}{,}{,}
\date{Dated: January 13, 2012}
\title{Early Warning with Calibrated and Sharper Probabilistic Forecasts}
\author{Reason L. Machete$^\dag$\\
{\small \dag Department of Mathematics and Statistics, University of Reading, United Kingdom}\\
{\small r.l.machete@reading.ac.uk}}
\begin{document}
\maketitle
\begin{abstract}
Given a nonlinear model, a probabilistic forecast may be obtained by Monte Carlo simulations. At a given forecast horizon, Monte Carlo simulations yield sets of discrete forecasts, which can be converted to density forecasts. The resulting density forecasts will inevitably be downgraded by model mis-specification. In order to enhance the quality of the density forecasts, one can mix them with the unconditional density. This paper examines the value of combining conditional density forecasts with the unconditional density. The findings have positive implications for issuing early warnings in different disciplines including economics and meteorology, but UK inflation forecasts are considered as an example.
\end{abstract}
{\renewcommand{\thefootnote}{}
\footnotetext{Abbreviations: Probability Integral Transform (PIT), Bank of England (BOE), Finite Unconditional Forecaster (FUF)}}
{\bf Keywords}:
{\small Calibration; Density forecasts; Combining forecasts; Scoring rule}
\section{Introduction}
Forecasts of a given quantity of interest often come from multiple sources. For instance, UK inflation forecasts are issued by both the Bank of England (BOE) and the National Institute of Economic and Social Research (NIESR)~\citep{hal-07}. In fact, the BOE has a suite of forecasting models which may be used to inform policy~\citep{kap-08}. Instead of determining the best model from multiple sources, it may be better to combine the models in some way~\citep{bun-89,cle-89,gra-ram}; but combining forecasts raises questions about a suitable criterion for assessing the quality of the composite forecast.

Combining forecasts has generally been viewed as a way of pooling information sources together~\citep{bat-69,gra-ram,bun-89,wal-05}.  Recently, model mis-specification has been given as another reason why pooling forecasts may be necessary~\citep{hend-04,kap-08}. \cite{hend-04} went further to suggest that pooling forecasts together may be viewed as a way of applying the James-Stein `shrinkage' estimation~\citep{jam-61}. Founded on multiple parameter estimation problems~\citep{jam-61,cas-85}, shrinkage estimates are obtained by mixing the maximum likelihood estimates with the `grand average'~\citep{efr-77}. The grand average is an average of all available data (or estimates). Shrinkage estimators are readily applicable to combining point forecasts~\citep[e.g.][]{grei-91}. Moreover, point forecasts have dominated the discussion on forecast combination~\citep{wal-05}. 

Hitherto, a notable departure from point forecasts to combining density forecasts was a discussion by~\cite{hal-07}, who also attributed earlier success of combined forecasts to model mis-specification. Considering UK inflation forecasts, their study confirmed that combining density forecasts outperforms the individual forecasts. Sources of the forecasts they considered were the BOE and NIESR. They combined these models with what they called `time series density'~\citep{hal-07}. Their `time series density' was a probability density function estimated from available data. The time series density may be considered an estimate of the unconditional density.

Here, we examine the quality of mixture distributions of conditional density forecasts and the unconditional density in light of the goal of probabilistic forecasting set forth by~\cite{til-gne}. The goal involves two concepts; {\em calibration} and {\em sharpness}. Calibration is the statistical consistency between forecast probabilities and observed relative frequencies~\citep{brie,til-gne,gne-08} while sharpness is a measure of how concentrated probabilistic forecasts are and a property of the forecasts only~\citep{bross-53,gne-08,til-gne,wilk-06}. Calibration was also termed {\it validity} by \cite{bross-53} and {\it reliability} by \cite{sau-58}.  Currently it is commonly known as {\em calibration}~\citep[e.g. in][]{gne-08,law-06}, although the weather community also uses the term reliability. While much of the discussion on calibration of probabilistic forecasts has centred on categorical events, \cite{daw-84} is notable for proposing the use of {\it probability integral transforms} (PITs) to assess the calibration of density forecasts. A PIT is obtained by plugging an observation into the cumulative predictive distribution function. His proposed test included the additional condition that the PITs should be independent and identically distributed. \cite{die-98} then showed that if density forecasts coincide with the ideal forecasts, then the PITs are independent and identically uniformly distributed (iid $U[0,1]$). 

The proposal of \cite{die-98} to use PITs was motivated by their quest for a universally applicable approach to density forecast evaluation. They argued that scoring rules cannot rank incorrect density forecasts in a way that satisfies all users. Recent work~\citep{til-07,joc-06} has discussed essential properties for scoring rules, but it is still unclear whether such can yield a consistent ranking of non ideal forecasts. On the other hand, testing whether PITs are iid $U[0,1]$ is only sufficient to determine if the forecasts are ideal or not. It is of no value in providing a universal ranking of non ideal forecasts.

 If the iid condition is relaxed, then PITs can be uniform even when the forecasts are not ideal. \cite{til-gne} termed this scenario {\it probabilistic calibration}. They introduced two other modes of calibration: {\it exceedance calibration} and {\it marginal calibration}. Marginal calibration refers to the case when the time average of all predictive distributions is equal to that of ideal forecasts. Since the time average of ideal forecasts can be estimated from time series, marginal calibration can be empirically assessed. There is no empirical way of assessing exceedance calibration and we will defer its definition until section~\ref{sub:cal}.

\cite{til-gne} then conjectured that when a subset of these modes of calibration holds, then the predictive distributions are at least as spread out as the ideal forecasts, which conjecture they termed a {\it sharpness principle}. The aim was to ensure that predictive distributions were no more confident than the ideal forecasts. It has further been argued that the goal of probabilistic forecasting is to maximise sharpness subject to calibration~\citep{gne-08,til-gne}. This so called {\it paradigm}~\citep{gne-08} depends on the aforementioned conjecture, which we shall revisit later. If one could identify relevant modes of calibration for the conjecture to hold, then the sharpness of predictive distributions could be maximised subject to those modes to achieve the goal. Maximising sharpness is equivalent to minimising uncertainty. 

This paper presents a new theoretical analysis of the quality of density forecasts in terms of sharpness and calibration. It focuses upon combining conditional forecasts with an unconditional estimate; this may be viewed as `shrinkage' of conditional forecasts towards the unconditional distribution~\citep{hend-04}. \cite{hal-07} found including the unconditional distribution to improve predictive distributions; but merely including the unconditional distribution cannot improve sharpness. Therefore, in addition to the mixture parameter, we suggest scaling the dispersion of conditional forecasts. Our analysis answers~\cite{hal-07}'s appeal for more theoretical work on combining density forecasts. Empirical results are given on UK inflation forecasts using the same example considered by~\cite{hal-07}.

The next section discusses forecast qualities that are cumulatively measured by the logarithmic scoring rule. In particular, a decomposition of this scoring rule is presented. The sharpness principle conjectured by \cite{til-gne} is discussed in \S~\ref{sec:spew} and a relevant proposition presented. In \S~\ref{sec:ens}, the methodology employed to produce density forecasts and theoretical analyses of forecast combinations are given. Results concerning density forecasts obtained via the logarithmic scoring rule with respect to the BOE inflation forecasts are presented in \S~\ref{sec:res}. Section~\ref{sec:con} gives a discussion and concluding remarks. Appendices~\ref{app:cal} and~\ref{app:sha} contain the proof of the proposition concerning the sharpness principle and appendix~\ref{app:proofs} proofs for the rest of the propositions. Appendix~\ref{app:point} gives a complementary discussion of point forecasts. 
\section{Probabilistic-Forecast Quality}
\label{sec:prob}
Model mis-specification places limitations on the value of probabilistic forecasts. On the other hand, consumers of forecasts may demand predictive distributions that are both {\it calibrated} and {\it sharp}. If such forecasts are issued at long time horizons, then early warning is afforded. We suggest that these qualities can be cumulatively  quantified by the logarithmic scoring rule proposed by~\cite{good-52}. There are other scoring rules available for selection~\citep[see][]{til-07}. For instance, there is the Brier score~\citep{brie}. This, however, decomposes into many terms~\citep{mur-93}, some of which are not relevant to our discussion and it is suitable for categorical events. A generalisation of the Brier score to density forecasts is the {\it continuous rank probability score}~\citep{til-07}, but it lacks a clear interpretation. Indeed traditional decompositions of scoring rules do not contain a sharpness term.  There is also the mean square error loss function~\citep{gra-06}, which is also irrelevant to the qualities of interest. Suffice it to say, the logarithmic scoring rule is preferred over others for its appeal to information theory concepts~\citep[see][]{roul-02}, which can be traced back to~\cite{sha,sha-49}. Information theory has a strong hold on uncertainty, a concept equivalent to sharpness.
\subsection{Logarithmic Scoring Rule}
Consider a density forecast $f(x)$ and a target probability density function $g(x)$. If we think of $X$ as a random variable, then the foregoing notation says that the true distribution of $X$ is $g(x)$. With this notation, the information based scoring rule used in this paper is
\begin{equation}
\mathbb{E}[\mbox{IGN}(f,X)]=-\int_{-\infty}^{\infty}g(x)\log f(x)\ud x,
\label{pro:eq1}
\end{equation}
where $\mbox{IGN}(f,X)=-\log f(X)$, proposed by~\cite{good-52} and termed {\em Ignorance} in~\cite{roul-02} and {\it predictive deviance} in \cite{knor-01}. Hence,~(\ref{pro:eq1}) is the expected Ignorance. It is related to the Kullback-Leibler divergence~\citep{kul-lei},
\[
D_{\mbox{KL}}(g||f)=\int_{-\infty}^{\infty}g(x)\log\left(\frac{g(x)}{f(x)}\right)\ud x
\]
by
\begin{equation*}
D_{\mbox{KL}}(g||f)=\mathbb{E}[\mbox{IGN}(f,X)]+\int_{-\infty}^{\infty}g(x)\log g(x)\ud x.
\end{equation*}
It follows that the $f$ that minimises $D_{\mbox{KL}}(g||f)$ also minimises $\mathbb{E}[\mbox{IGN}(f,X)]$. The expected Ignorance is the cross entropy $H(g,f)$. The Ignorance score is especially relevant when one evaluates the performance of density forecasts given time series only, with no access to $g(x)$. An important property of the Ignorance score is that it attains the minimum if and only if $f(x)=g(x)$~\citep{joc-07,til-07}, meaning it is {\it strictly proper}.

Traditionally, the only score that has been decomposed into constituent terms is the Brier score: the {\it reliability-resolution} decomposition~\citep{mur-93,wilk-06}, after removing the uncertainty term. \cite{joc-09} extended the decomposition to general scores, but in the context of categorical forecasts. Unlike sharpness, resolution is not a property of the forecasts only. Therefore, we introduce a decomposition of~(\ref{pro:eq1}) as
\begin{equation*}
\mathbb{E}[\mbox{IGN}(f,X)]=-\int_{-\infty}^{\infty}f(x)\log f(x)\ud x-\int_{-\infty}^{\infty}\left[g(x)-f(x)\right]\log f(x)\ud x.
\end{equation*}
In this decomposition of expected Ignorance, the first term is {\it sharpness} and the second is {\it calibration}. Notice that the sharpness term is simply the density entropy $H(f)$, a property of the density forecast only. It is desirable for this term to be as negative as possible, effectively expressing more certainty about what is likely to happen. Since calibration is a statistical property of the forecasting system, it cannot be assessed based on one forecast only. For a time series of forecasts, we want each $f(x)$ to be close to $g(x)$ in some way. One is never furnished with $g(x)$ to aid assessment of calibration in an operational setup, but there are time series approaches to address this.
\subsection{Sharpness}
One way to quantify sharpness is to use the variance~\citep[e.g.][]{til-gne}. We emphasise that sharpness should be quantified by entropy, which ``is a measure of concentration'' of the distribution ``on a set of small measure'', a small value of entropy corresponding to a ``high degree of concentration''~\citep{hir-57}. The entropy of a distribution $f(x)$ of variance $\sigma^2$ satisfies the inequality~\citep{sha,sha-49}
\begin{equation*}
-\int_{-\infty}^{\infty}f(x)\log f(x)\ud x\le\frac{1}{2}\log\left(2\pi e\sigma^2\right).
\end{equation*}
Hence, a smaller variance guarantees lower entropy but not vice versa. Indeed two distributions with the same variances can have unequal entropies. For instance, a mixture of two Gaussians will have lower entropy than a single Gaussian distribution of the same variance. Much more, a distribution of a higher variance can have a lower entropy than that of lower variance.

Sharpness has also been quantified by confidence intervals~\citep{raf-05,til-gne}. Confidence intervals share a similar weakness to variance in the sense that a bimodal distribution that is fairly concentrated on the two modes can have larger confidence intervals than a unimodal distribution that is fairly spread out. Also, given two non-symmetric distributions, which of them is deemed sharper could depend on what the confidence level is. 
\subsection{Calibration}
\label{sub:cal}
The calibration of density forecasts is a well trodden subject. Much of the literature takes the stand that a calibrated forecasting system is tantamount to a correctly specified model. \cite{gra-06} provide a comprehensive survey of formal statistical techniques for assessing calibration of density forecasts to determine if the underlying model is correctly specified. The work of \cite{til-gne} strikes a discord by providing a calibration framework that accommodates model mis-specification. They broke down calibration into three modes, each of which could be assessed separately.

Suppose a probability forecasting system issues predictive distributions $\{F_t(x)\}_{t=1}^T$, while the data-generating process issues ideal forecasts $\{G_t(x)\}_{t=1}^T$. \cite{til-gne} then defined the following modes of calibration: 
\begin{itemize}
\item The sequence $\{F_t(x)\}_{t=1}^T$ is {\em probabilistically calibrated} relative to $\{G_t(x)\}_{t=1}^T$ if
\begin{equation}
\frac{1}{T}\sum_{t=1}^TG_t\{F_t^{-1}(p)\}=p,\quad p\in(0,1).
\label{eqn:prob}
\end{equation}
\item The sequence $\{F_t(x)\}_{t=1}^T$ is {\em exceedance calibrated} relative to $\{G_t(x)\}_{t=1}^T$ if 
\begin{equation}
\frac{1}{T}\sum_{t=1}^TG_t^{-1}\{F_t(x)\}=x,\quad x\in \Re.
\label{eqn:exc}
\end{equation}
\item The forecaster is {\em marginally calibrated} if
\begin{equation*}
\lim_{T\rightarrow\infty}\frac{1}{T}\sum_{t=1}^TF_t(x)=\lim_{T\rightarrow\infty}\frac{1}{T}\sum_{t=1}^TG_t(x).
\end{equation*}
\end{itemize}
Note that the definitions of probabilistic and exceedance calibration require the distributions to be strictly increasing over all $\Re$. In the subsequent discussions, we admit cases where the distributions simply have compact support. In such cases, the inverses will only be taken within the regions of compact support. If we have a time series of observations $x_t$, then $z_t=F_t(x_t)$ is a {\em probability integral transform} (PIT) \citep{gra-06,die-98}. Uniformity of the PITs is equivalent to probabilistic calibration~\citep{til-gne}. A visual inspection of PIT histograms would reveal obvious departures from uniformity. The underlying model is correctly specified if and only if $z_t\sim\mbox{iid}$ $U[0,1]$.

Suppose we have a time series of density forecasts, $\{f_t(x)\}_{t\ge1}$. Then define the {\em forecaster's unconditional density} as
\begin{equation*}
\tilde{\rho}_T(x)=\frac{1}{T}\sum_{t=1}^Tf_t(x).
\end{equation*}
We define a forecaster who issues the finite time unconditional distribution,
\begin{equation*}
F_t(x)=\bar{G}_T(x)=\frac{1}{T}\sum_{t=1}^TG_t(x)
\end{equation*} 
for all $t\in\{1\ldots,T\}$, to be the {\it finite unconditional forecaster} (FUF). If $F_t(x)=\lim_{T\rightarrow\infty}\bar{G}_T(x)$, then we have the {\it unconditional forecaster}\footnote{\cite{til-gne} refer to this as the {\em climatological forecaster}, even though this may have nothing to do with climate in a meteorological sense.}. A forecaster is {\it finite marginally calibrated} if $\tilde{\rho}_T(x)=\bar{G}_T'(x)$.

For all practical purposes, $T$ is finite and we have no access to the $G_t(x)$'s. Hence it is difficult to assess finite marginal calibration. If $d(\tilde{\rho}_T,\tilde{\rho}_{2T})\approx 0$, where $d$ is some metric, then we can take $T$ to be large enough to evaluate marginal calibration. To this end, we can use the Hellinger distance~\citep{pol-02} and compute
\begin{equation*}
h(\tilde{\rho}_T,\rho_u)=\frac{1}{2}\int\left[\sqrt{\tilde{\rho}_T(x)}-\sqrt{\rho_u(x)}\right]^2\ud x,
\end{equation*}
where $\rho_u(x)=\lim_{T\rightarrow\infty}\bar{G}_T'(x)$ is the underlying system's {\em unconditional density}. It is useful to note that $0\le h(\cdot,\cdot)\le 1$, assuming the value of 0 when the two distributions are identical and 1 when they do not overlap. This procedure for assessing marginal calibration is an alternative to the graphical tests performed in~\cite{til-gne}. It is expected to be more robust to finite sample~effects.
\section{The Sharpness Principle and Early Warning}
\label{sec:spew}
\cite{mur-98} highlighted that forecasts need to be calibrated before one worries about sharpness. Recently, \cite{til-gne} adopted a paradigm of maximising sharpness subject to calibration. They then conjectured that the goal to obtain ideal forecasts and of maximising sharpness subject to calibration are equivalent, which is the {\em sharpness principle}. To revisit the conjecture, denote the variance of a random variable whose distribution is $F(x)$ by $\mbox{var}(F)$. Then \cite{til-gne} define a forecaster to be at least as spread out as the ideal forecaster if the inequality 
\begin{equation}
\frac{1}{T}\sum_{t=1}^T\mbox{var}(F_t)\ge\frac{1}{T}\sum_{t=1}^T\mbox{var}(G_t)
\label{eqn:var}
\end{equation}
holds. With this notion of spread, a weaker alternative states that any sufficiently calibrated forecaster is at least as spread out as the ideal forecaster \citep{til-gne}. The key word here is ``sufficiently''. Maintaining our reservations about using variance to quantify sharpness, this section is concerned with addressing this weaker conjecture.
\subsection{Counter Examples}
None of the individual modes of calibration alone is sufficient for the weaker conjecture to hold \citep{til-gne}. In this subsection, we present relevant counter examples. 

{\bf Probabilistically calibrated forecaster}: Let $G_t=U[0,1]$, $t=1,2$, be ideal forecasts and the corresponding forecaster 
\begin{equation*}
F_t(x)=\left\{\begin{array}{cc}
0, & x<0,\\
x/\{2\beta^{2-t}(1-\beta)^{t-1}\},& x\in[0,\beta],\\
\frac{1}{2}+(x-\beta^{2-t}(1-\beta)^{t-1})/\{2\beta^{t-1}(1-\beta)^{2-t}\}, &x\in[\beta,1],\\
1, & x\ge1
\end{array}\right.
\end{equation*}
for $t=1,2$ with $0<\beta<1/2$. This forecaster satisfies equation~(\ref{eqn:prob}), hence is probabilistically calibrated. Note that when $\beta=1/3$, the average variance of the forecaster distributions is 37/432 whilst the ideal forecaster yields 1/12. Hence, inequality~(\ref{eqn:var}) is satisfied. If entropy is used to measure sharpness, then $H(f_t)=(1/2)\log(4\beta(1-\beta))$ for $t=1,2$, where $H(f_t)$ denotes the density entropy of $f_t$ and $f_t(x)=F_t'(x)$. Since $H(g_t)=0$ and $4\beta(1-\beta)<1$, $H(f_t)<H(g_t)$. Thus entropy indicates that the forecaster is sharper than the ideal forecaster. For distributions with the same compact support, it is well known in information theory that the uniform distribution yields maximum uncertainty, a fact that is missed by using variance.

\cite{pal-10} gives an example in which~(\ref{eqn:var}) is violated. He takes
$$G_1(x)=1-2e^{-\lambda x}+e^{-2\lambda x},\quad G_2(x)=1-e^{-2\theta x},\quad F_1(x)=1-e^{-\lambda x},\quad F_2(x)=1-e^{-\theta x},$$
supported on $(0,\infty)$, with $\theta^2>3\lambda^3$. Here, $\mbox{var}(F_1)=1/\lambda^2$, $\mbox{var}(F_2)=1/\theta^2$, $\mbox{var}(G_1)=5/(4\lambda^2)$ and $\mbox{var}(G_2)=1/(4\theta^2).$ Hence~(\ref{eqn:var}) does not hold, yet the forecaster is probabilistically calibrated.

{\bf Exceedance calibrated forecaster}: Now let $G_t=U[0,1]$, $t=1,2$ be the ideal forecaster whose corresponding forecaster is:
$$F_t=U\left[\frac{t-1}{2},\frac{t}{2}\right],$$
$t=1,2$. This forecaster is exceedance calibrated since, 
\begin{eqnarray*}
\frac{1}{2}\left[G_1^{-1}\{F_1(x)\}+G_2^{-1}\{F_2(x)\}\right]&=&\frac{1}{2}\left\{\begin{array}{ll}
2x+0, & x\in[0,1/2),\\
1+2(x-1/2), & x\in[1/2,1],\\
\end{array}\right.\\
&=&\begin{array}{ll}
x, & x\in(0,1)
\end{array}
\end{eqnarray*}
yet~(\ref{eqn:var}) is violated because the average variance of the forecaster is $1/48$ $(<1/12)$. We also have $H(f_t)=-\log 2$~$(<H(g_t))$. Both measures concur that the forecaster is sharper than the ideal forecaster.

{\bf Marginally calibrated forecaster}: Suppose $G_t=[0,1]$, $t=1,2,\ldots,\infty$ be a sequence of ideal forecasts and suppose that a forecaster issues
$$F_t=U\left[(k_t-1)/n,k_t/n\right]$$
for some finite $n>1$ and $k_t$ is a discrete uniform random variable taking values $\{1,2,..,n\}$\footnote{Prof T. Gneiting brought this example to my attention through private communication.}. Clearly, the unconditional distribution is $U[0,1]$. The forecaster's unconditional distribution is given by
\begin{equation*}
F(x)=\lim_{T\rightarrow\infty}\frac{1}{T}\sum_{t=1}^TF_t(x).
\end{equation*}
Note that $F_t(x)$ is a function of a random variable due to its dependence on $k_t$. It turns out that the expectation of $F_t$ is $U[0,1]$. Hence, by the law of large numbers, $F=U[0,1]$. Hence the forecaster issuing $F_t$ is marginally calibrated. However, both entropy and variance indicate that the forecaster is sharper than the ideal forecaster.
\subsection{Calibration Theorem}
Since none of the individual modes of calibration is sufficient for the forecaster to be less sharp than the ideal forecaster~\cite{til-gne}, we ought to determine if any two would suffice. Since \cite{pal-09} did not satisfactorily address this~\citep{pal-10}, it is revisited. In order to address this conjecture, we define {\em finite marginal calibration} as 
\begin{equation}
\frac{1}{T}\sum_{t=1}^TF_t(x)=\frac{1}{T}\sum_{t=1}^TG_t(x).
\label{eqn:mar}
\end{equation}
From a practical point of view, probabilistic and marginal calibration are more important than exceedance calibration because they can be assessed empirically.
\theoremstyle{plain} \newtheorem{proposition}{PROPOSITION}
\begin{proposition}
\label{sharpness}
Suppose $\{G_t\}_{t=1}^T$ is a sequence of continuous and strictly increasing distribution functions (ideal forecasts). Then a forecaster who is both probabilistically and finite marginally calibrated has either issued ideal forecasts $\{G_t\}_{t=1}^T$ or is the finite unconditional forecaster.
\end{proposition}
The proof for the above proposition is split into two parts and is given in appendix~\ref{app:cal} and \ref{app:sha}. It is trivial that the ideal forecaster satisfies the probabilistic calibration condition~(\ref{eqn:prob}) and the finite marginal calibration condition~(\ref{eqn:mar}). It is also trivial that the finite unconditional forecaster (FUF) satisfies the finite marginal calibration condition. To show that the FUF is probabilistically calibrated, note that for any $p\in(0,1)$, there exists $\eta$ such that $\bar{G}_T(\eta)=p\Rightarrow\bar{G}_T^{-1}(p)=\eta$. Hence
\begin{eqnarray*}
p&=&\overline{G}_T(\eta)\\
 &=&\frac{1}{T}\sum_{t=1}^TG_t(\eta)\\
&=&\frac{1}{T}\sum_{t=1}^TG_t\{\bar{G}_T^{-1}(p)\}.
\end{eqnarray*}

Including exceedance calibration in the hypotheses of the above proposition would rule out the FUF. If the FUF was exceedance calibrated, we would have 
\begin{eqnarray*}
\frac{1}{T}\sum_{t=1}^TG_t^{-1}\{\bar{G}_T(x)\}=x.
\end{eqnarray*}
Given an $x$, there exists a $\xi\in(0,1)$ such that $\bar{G}_T(x)=\xi\Rightarrow x=\bar{G}_T^{-1}(\xi)$. Hence, we can eliminate $x$ in the above equality to obtain
\begin{eqnarray*}
\frac{1}{T}\sum_{t=1}^TG_t^{-1}(\xi)&=&\bar{G}_T^{-1}(\xi)\\
&=&\left(\frac{1}{T}\sum_{t=1}^TG_t\right)^{-1}(\xi).
\end{eqnarray*}
Unless $G_t(x)=G(x)$ for all $t$, the above equality is a mathematical fallacy. Therefore, the FUF cannot be exceedance calibrated.

Even though this proposition does not deal with the case when $T$ approaches infinity, in all practical situations we deal with finite $T$. Indeed the graphical tests for marginal calibration discussed in \cite{til-gne} deal with {\it finite} marginal calibration. Defining the average predictive distribution to be
\begin{equation*}
\bar{F}_T(x)=\frac{1}{T}\sum_{t=1}^TF_t(x)
\end{equation*}
and the empirical CDF of the observations by
\begin{equation*}
\hat{G}_T(x)=\frac{1}{T}\sum_{t=1}^T{\bf 1}(x_t<x), 
\end{equation*}
where $x_t$ is time series, \cite{til-gne} propose plotting a graph of $\bar{F}_T-\hat{G}_T$ against $x$ to assess marginal calibration. Clearly this is assessing finite marginal calibration.

The implications (of the proposition) to the goal of probabilistic forecasting are that the level of expectation with regard to the two modes of calibration needs to be scaled down when the underlying model is mis-specified. This is because, without a correctly specified model, one cannot have both perfect probabilistic and finite marginal calibration unless he is the FUF. Hence the forecaster should merely aim to maximise sharpness subject to some level of calibration. For given levels of probabilistic and finite marginal calibration, a forecaster affords early warning if he is sharper than the unconditional distribution.
\section{Density-Forecast Estimation}
\label{sec:ens}
Suppose we have some data point $s_t$, at time $t$, and we want to know the future state at time $t+\tau$. We call $\tau$ the {\it forecast lead time}. In order to express uncertainty in the forecasts, we issue a density forecast. One way to obtaining a density forecast is to generate many points in the neighbourhood of $s_t$ and iterate them forward with the model to obtain an ensemble of forecasts $\boldsymbol{X}^{(t+\tau)}=\left\{X_i^{(t+\tau)}\right\}_{i=1}^N$ at time $t+\tau$. A comprehensive review of how to generate ensemble forecasts with non-linear models is given by~\cite{leu-08}, but if we think of this process as Monte Carlo simulation, we may refer to~\cite{clem-01} for a brief description in an econometric setting. This section is concerned with converting the ensemble into a density forecast. The Gaussian kernel function,
\begin{equation*}
K(\xi)=\frac{1}{\sqrt{2\pi}}\exp\left(-\xi^2/2\right),
\end{equation*} 
may be used to obtain a density function from an ensemble of point forecasts.
\subsection{Single Model}
One way to convert a forecast ensemble into a density forecast would be to perform density estimation according to~\cite{par-62} and \cite{sil-86}. The fundamental weakness of this approach is that it inherently assumes that the ensemble is a draw from the true distribution. In view of this, \cite{roul-02} suggested taking into account how the model has performed in the past. A similar approach is followed by~\cite{hal-07}, who use past forecast errors to obtain density forecasts. Therefore, we can form density forecast estimates of the form:
\begin{equation}
\rho^{(t)}(x)=\frac{1}{\sigma N}\sum_{i=1}^NK\left\{\left(x-X_i^{(t)}-\mu\right)/\sigma\right\},
\label{eqn:ei1}
\end{equation}
where $\sigma$ and $\mu$  are respective kernel width and offset parameters chosen according to past performance and $K(\cdot)$  is the kernel function. The density forecast in~(\ref{eqn:ei1}) differs from the traditional~\cite{par-62} estimates by the offset parameter. It is similar to the Bayesian Model Average proposed by~\cite{raf-05} with a uniform bias correction, $\mu$ and equal weights. Selecting $\sigma$ using \cite{sil-86} does not account for model mis-specification.

To account for model mis-specification, let us first denote a record of past time series and corresponding ensemble forecasts by $\mathcal{V}_T=\left\{\right(s_t,\boldsymbol{X}^{(t)})\}_{t=1}^T$. Then the density forecasts whose parameters, $\mu$ and $\sigma$, are selected by taking into account past performance may be denoted by $\rho^{(t)}(x|\mathcal{V}_T)$. While $\rho^{(t)}(x|\mathcal{V}_T)$ has the same form as in~(\ref{eqn:ei1}), its parameters are selected by doing the minimisation
\begin{equation}
\min_{\sigma>0,\mu}\left\{-\frac{1}{T}\sum_{t=1}^T\log\rho^{(t)}(s_t|\mathcal{V}_T)\right\}.
\label{eqn:ei4}
\end{equation}
Under certain assumptions, doing the minimisation in~(\ref{eqn:ei4}) is tantamount to minimising either the average cross entropy or the average Kullback-Leibler divergence. Without making any assumptions, the term in~(\ref{eqn:ei4}) should be called average Ignorance, $\langle \mbox{IGN}\rangle$. Minimising~(\ref{eqn:ei4}) is equivalent to maximum likelihood under the assumption of independence of forecast errors~\citep{raf-05}. Moreover, it is equivalent to {\it quasi maximum likelihood} (QML) under model mis-specification with independent conditional forecasts as discussed by~\cite{whi-94}. Interestingly, \cite{whi-82} called the QML estimator the `minimum ignorance' estimator, arguing that it minimises our ignorance about the correct model structure.
\subsection{Mixture Model}
\cite{joc-07} noted that, when doing the minimisation in~(\ref{eqn:ei4}), some of the $\boldsymbol{X}^{(t)}$ may be far from the corresponding $s_t$, which could result in choices of $\sigma$ that were too big. Hence, the parameter estimates would not be robust. These short comings could largely be due to model mis-specification. To circumvent these, they proposed a mixture model of the unconditional density, $\rho_u(x)$, and $\rho^{(t)}(x|\mathcal{V}_T)$:
 \begin{equation}
f^{(t)}(x|\mathcal{V}_T)=\alpha\rho^{(t)}(x|\mathcal{V}_T)+(1-\alpha)\rho_u(x),
\label{eqn:ei2}
\end{equation}
where the mixture parameter, $\alpha\in[0,1]$. All the three parameters are fitted simultaneously by minimising average Ignorance. The unconditional density, $\rho_u(x)$, is estimated from data via
\begin{equation*}
\rho_u(x)=\frac{1}{\sigma_u T}\sum_{t=1}^TK\left\{\left(x-s_t-\mu_u\right)/\sigma_u\right\},
\end{equation*}
and the parameters $\sigma_u$ and $\mu_u$ are then chosen to simultaneously minimise the logarithmic scoring rule as proposed in \cite{joc-07}. \cite{sil-86} may also be followed to estimate the unconditional density.

Note that~(\ref{eqn:ei2}) would be the {\em linear opinion pool} discussed in~\cite{cle-99} and used in~\cite{hal-07} if $\alpha$ was the only parameter being selected; but we also train the $\sigma$ to enhance the sharpness of the mixture distribution~(see Proposition~\ref{prop:sharp}). The role of $\alpha$ is like that of the shrinkage in multi-parameter estimation~\citep{efr-77,hend-04}. If we let $r_t=\rho^{(t)}(s_t)/\rho_u(s_t)$, then we can state the following proposition:
\begin{proposition}
\label{prop:returns}
For a given set of parameters $\mu$ and $\sigma$, the necessary and sufficient conditions for improvement from including the unconditional density in the sense of the logarithmic scoring rule are that
\begin{equation*}
\frac{1}{T}\sum_{t=1}^Tr_t>1\quad\mbox{and}\quad\frac{1}{T}\sum_{t=1}^T\frac{1}{r_t}>1.
\end{equation*}
\end{proposition}
The proof for this proposition is given in appendix~\ref{app:proofs}. Its counter part for point forecasts is Proposition~\ref{prop:point} in Appendix~\ref{app:point}. The ratio $r_t$ may be interpreted as the {\it return ratio} on some invested capital in a Kelly betting scenario~\citep{kel-56} with no track take. The proposition states how the conditional and unconditional densities are to outperform each other in order for the mixture model to provide additional value.

In order to capture the effect, on kernel width, of including the unconditional density, we consider the case when $N=1$ with $\mu=0$. When there is no unconditional density included, minimising the logarithmic score yields,
\begin{equation}
\sigma_o^2=\frac{1}{T}\sum_{t=1}^T\left\{s_t-X_1^{(t)}\right\}^2.
\label{eqn:ei7}
\end{equation}
Let us write a time series version of the logarithmic scoring rule as
\begin{equation}
\langle\mbox{IGN}\rangle=-\frac{1}{T}\sum_{t=1}^T\log f^{(t)}(s_t|\mathcal{V}_T).
\label{eqn:ei5}
\end{equation}
\begin{proposition}
\label{prop:discount}
Suppose the score given by equation~(\ref{eqn:ei5}) assumes a minimum at parameter values $(\sigma_*,\alpha_*)$, then the following equation holds:
\begin{equation}
\sigma_*^2=\frac{1}{T}\sum_{t=1}^T\left\{s_t-X_1^{(t)}\right\}^2\frac{\rho^{(t)}(s_t|\mathcal{V}_T)}{f^{(t)}(s_t|\mathcal{V}_T)}.
\label{eqn:ei6}
\end{equation}
\end{proposition}
See appendix~\ref{app:proofs} for the proof. Corollary~\ref{app:cor} in Appendix~\ref{app:point} gives corresponding sharpness conditions for point forecasts. For illustrative purposes, suppose that the $k$th forecast is far from the corresponding observation in the sense that $$\left|s_k-X_1^{(k)}\right|\gg\max\left\{\left|s_t-X_1^{(t)}\right|\right\}_{t\neq k}.$$
As a result, the kernel width in (\ref{eqn:ei7}) would be inflated. Equation~(\ref{eqn:ei6}) provides a way to discount the contributions of a few bad forecasts on the kernel width. In this case, $(\sigma_*,\alpha_*)$ would be chosen such that 
$$\frac{\rho^{(k)}(s_k|\mathcal{V}_T)}{f^{(k)}(s_k|\mathcal{V}_T)}\ll 1.$$
This is especially valuable when $T$ is small, which is the case in typical time series. The idea is that a reduction in kernel width is necessary for the entropy of $f^{(t)}(x|\mathcal{V}_T)$ to decrease even when $N>1$, but it is easier to explain how the reduction is achieved when $N=1$. Despite this reduction, some mixture forecasts may still be less sharp than unconditional distribution in the sense of entropy. A straight forward application of the Kullback-Leibler (KL) and Jensen's inequalities leads to the relations
\begin{eqnarray}
\nonumber
\alpha H\left\{\rho^{(t)}\right\}+(1-\alpha)H(\rho_u)\le H\left\{f^{(t)}\right\}\le\alpha^2H\left\{\rho^{(t)}\right\}+\alpha(1-\alpha)H\left\{\rho^{(t)},\rho_u\right\}+\ldots\\(1-\alpha)\alpha H\left\{\rho_u,\rho^{(t)}\right\}+(1-\alpha)^2H(\rho_u),
\label{eqn:i9}
\end{eqnarray}
where $H(f)=-\int f(x)\log f(x) \ud x$ and $H(f,g)=-\int f(x)\log g(x)\ud x$ are the entropy and cross entropy respectively. Therefore, the necessary and sufficient conditions for $ H\left\{f^{(t)}\right\}\ge H(\rho_u)$ to hold are that $H(\rho_u)<H\left\{\rho^{(t)},\rho_u\right\}$ and $H(\rho_u)<H\left\{\rho^{(t)}\right\}$ respectively. The first inequality of~(\ref{eqn:i9}) can be used to establish the following proposition:
\begin{proposition}
\label{prop:sharp}
If $H\{\rho^{(t)}\}<H(\rho_u)$, then $H\{\rho^{(t)}\}\le H\{f^{(t)}\}$, i.e. in the sense of entropy, merely mixing $\rho^{(t)}(x)$ with the unconditional density without re-adjusting the $\sigma$ parameter cannot improve the sharpness of the predictive distribution.
\end{proposition}
It is not obvious what the effect of the mixture is on calibration, except that including the unconditional density improves the KL distance from the ideal forecasts. Nevertheless, the mixture parameter that minimises the logarithmic score yields the equation
\begin{equation*}
\frac{1}{T}\sum_{t=1}^T\frac{\rho_c(s_t)}{f^{(t)}(s_t|\mathcal{V}_T)}=1.
\end{equation*}
On the other hand, we note that equation~(\ref{eqn:prob}) is equivalent to
\begin{equation}
\frac{1}{T}\sum_{t=1}^T\frac{g_t(s_t)}{f_t(s_t)}=1
\label{eqn:ei8}
\end{equation}
The two preceding equations are similar with the $\rho_u$ replacing $g_t$ in~(\ref{eqn:ei8}). What happens to calibration due the mixture will be explored by way of example in the next section.
\section{Applications}
\label{sec:res}
This section presents the results that highlight the effects, on sharpness, calibration and the time horizon over which density forecasts are useful, of introducing the unconditional density to form the density forecasts. As an example, the Bank of England (BOE) inflation forecasts are considered. Every quarter, the BOE issues GDP and inflation quarterly forecasts for up to twelve quarters into the future~\citep{harr-05}. The forecasts come in pairs; those based on constant interest rates and those on market interest rates. We shall only consider those based on constant interest rates. The model is nonlinear and is an addition to a suite of models which are admittedly imperfect since they are simplifications of reality~\citep{harr-05,kap-08}. 

We shall here consider the Retail Price Index inflation excluding mortgage interest payments (RPIX inflation rate). The corresponding forecasts are published on the BOE website for the period from 1993 to 2005. The starting point of this period coincided with the BOE starting to issue an inflation target of 2.5\%. Before that time there were no inflation targets set. RPIX inflation data is published on the Office of National Statistics website.

At a given lead time, each forecast comprises the parameters mode, mean, median, uncertainty and the skew parameter. A two-piece normal distribution is then used to produce fan-charts~\citep{bri-98} which are then published on the BOE website. Only the parameters mode (or central projection), $\mu_t$, uncertainty, $\sigma_t$, and skewness, $\gamma_t$, are required to completely specify the two-piece distribution. In a nutshell, the probability distribution issued is~\citep{bri-98,gne-11}
\begin{equation}
\rho_t(x;\mu_t,\sigma_{1,t},\sigma_{2,t})=\left\{\begin{array}{ll}
\left(\frac{2}{\pi}\right)^{1/2}(\sigma_{1,t}+\sigma_{2,t})^{-1}\exp\left\{-\frac{(x-\mu_t)^2}{2\sigma_{1,t}^2}\right\}, & x<\mu_t\\\\
\left(\frac{2}{\pi}\right)^{1/2}(\sigma_{1,t}+\sigma_{2,t})^{-1}\exp\left\{-\frac{(x-\mu_t)^2}{2\sigma_{2,t}^2}\right\}, & x\ge\mu_t,
\end{array}\right.
\end{equation}
where $$\sigma_{1,t}=\sigma_t/\sqrt{1+\gamma_t},\quad\sigma_{2,t}=\sigma_t/\sqrt{1-\gamma_t}.$$
One may think of the central projection as a one member ensemble (or Monte Carlo simulation). The entropy of this density function is given by
$$H(\rho_t)=\log\left\{\sqrt{\pi/2}(\sigma_{1,t}+\sigma_{2,t})\right\}+\frac{1}{2}.$$

\cite{dow-07} highlighted that the BOE RPIX inflation forecasts are very pessimistic by assessing the corresponding fan charts. He argued that the BOE over-estimated the probability that a given target range would be breached. In this paper, we demonstrate that the distributions can be made sharper by mixing them with the unconditional density. To this end, the new forecast density shall be given by
\begin{equation}
f_t(x)=\alpha\rho_t(x;\lambda\sigma_{1,t},\lambda\sigma_{2,t})+(1-\alpha)\rho_u(x).
\label{eqn:mix1}
\end{equation}
The parameters $\lambda$ and $\alpha$ are selected by minimising Ignorance over a forecast-verification archive and they depend on lead time. The use of $\lambda$ in~(\ref{eqn:mix1}) accommodates the BOE's judgement and allows the mixture distributions to be sharper than the BOE forecasts. Forecast combination as done in~\cite{hal-07} corresponds to setting $\lambda=1$ and then finding an `optimal' $\alpha$. According to Proposition~\ref{prop:sharp}, this cannot yield sharper forecasts. The case $\lambda=1$ and $\alpha=1$ corresponds to the BOE forecasts and in sequel, this situation shall simply be referred to as $\alpha=1$.

In order to estimate the unconditional density, we need to make sure that inflation exhibited stationarity within the epoch under investigation. The works of \cite{boe-08} and \cite{ben-04} argue that the inflation targeting policy of 1992 introduced a break in the dynamics of inflation and marked the beginning of a period of remarkable stability. Using unit-root tests, they found that giving independence to the BOE in 1997 introduced no break in the dynamics of RPIX inflation. In light of this, we estimated the unconditional density using data for the period from 1992 to 2004.

For each forecast lead time considered, there were 44 forecasts. Since this is a small data size to deal with, we perform a cross validation approach to quantify the effect of including the unconditional density on the sharpness and calibration of the density forecasts. This is done by successively leaving out one forecast from the training set. Each 43 member training set is then used to estimate values of $\lambda$ and $\alpha$ that minimise Ignorance. These values are then used on each excluded forecast to produce the mixture density forecast. Entropies and PITs of these out of sample distributions are then compared with those of the BOE forecasts. We also compare the out of sample forecaster's unconditional density with the BOE unconditional density via the Hellinger distance.
\begin{figure}[tp]
\centering
\hbox{
\epsfig{file=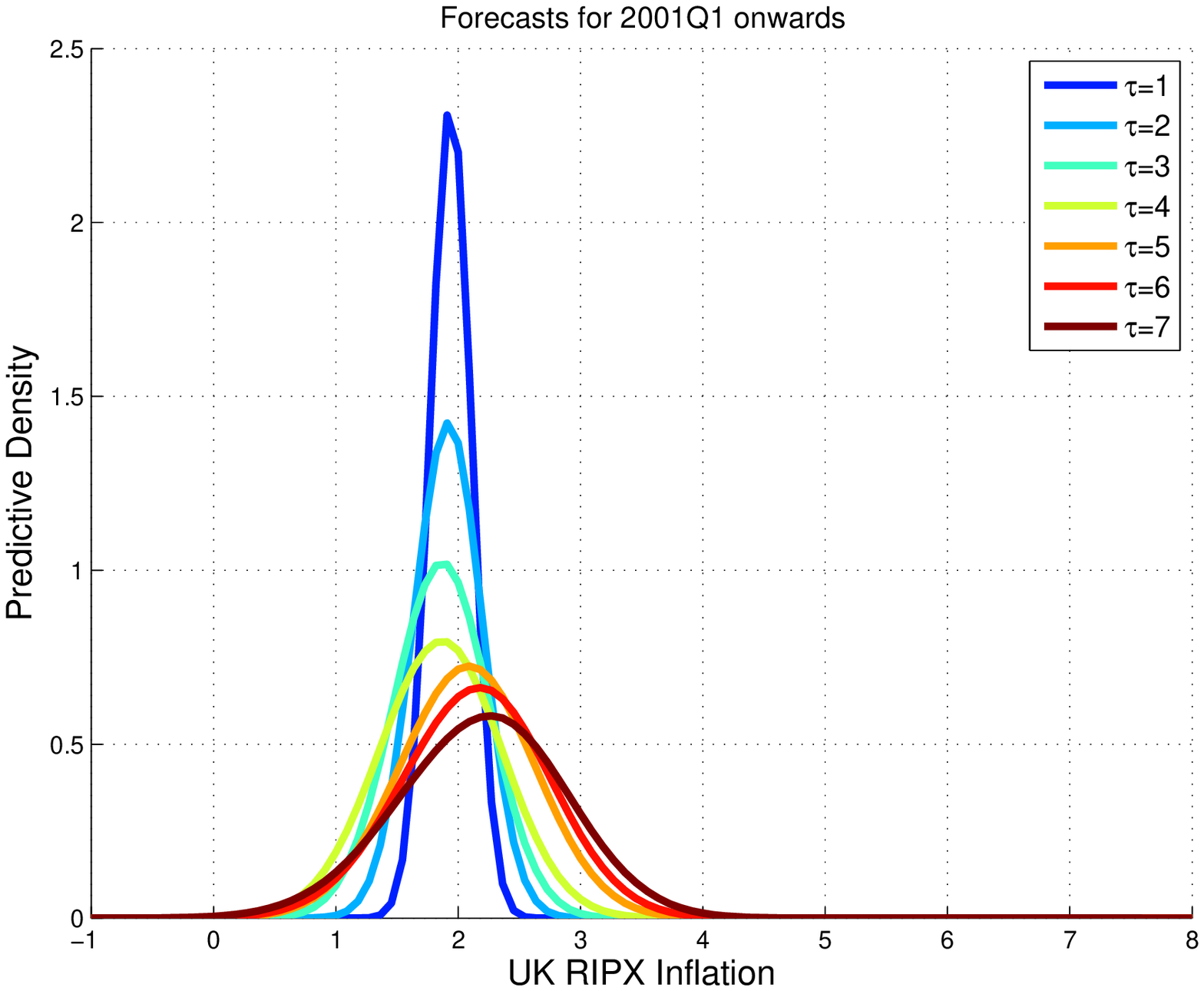,width=7.5cm,height=7.5cm},
\epsfig{file=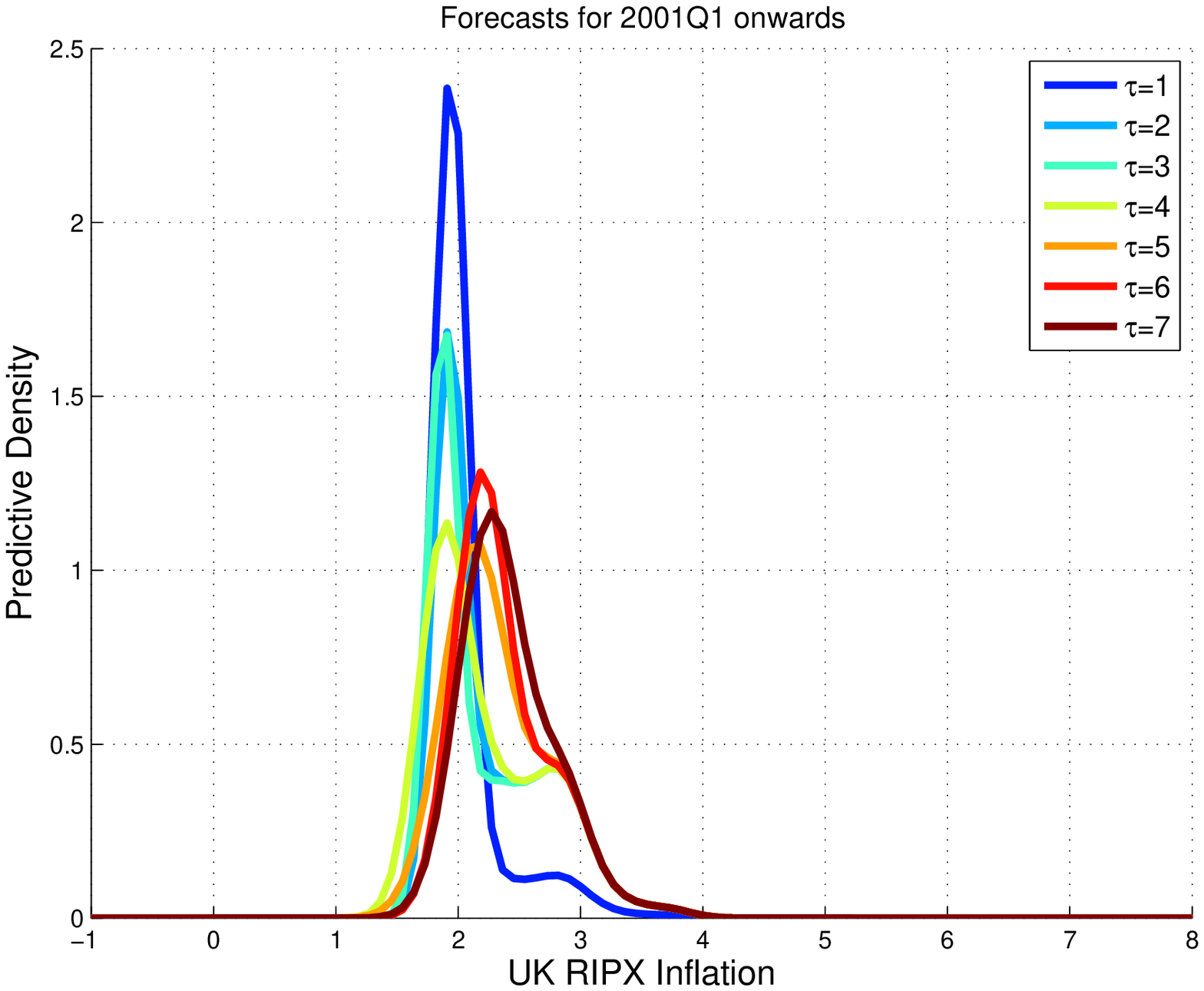,width=7.5cm,height=7.5cm}
}
\caption{\small\em Graphs of density forecasts of RPIX inflation as issued by the Bank of England (left) and when unconditional density is included (right).}
\label{res:rpdfs}
\end{figure}

\begin{figure}[tp]
\centering
\hbox{
\epsfig{file=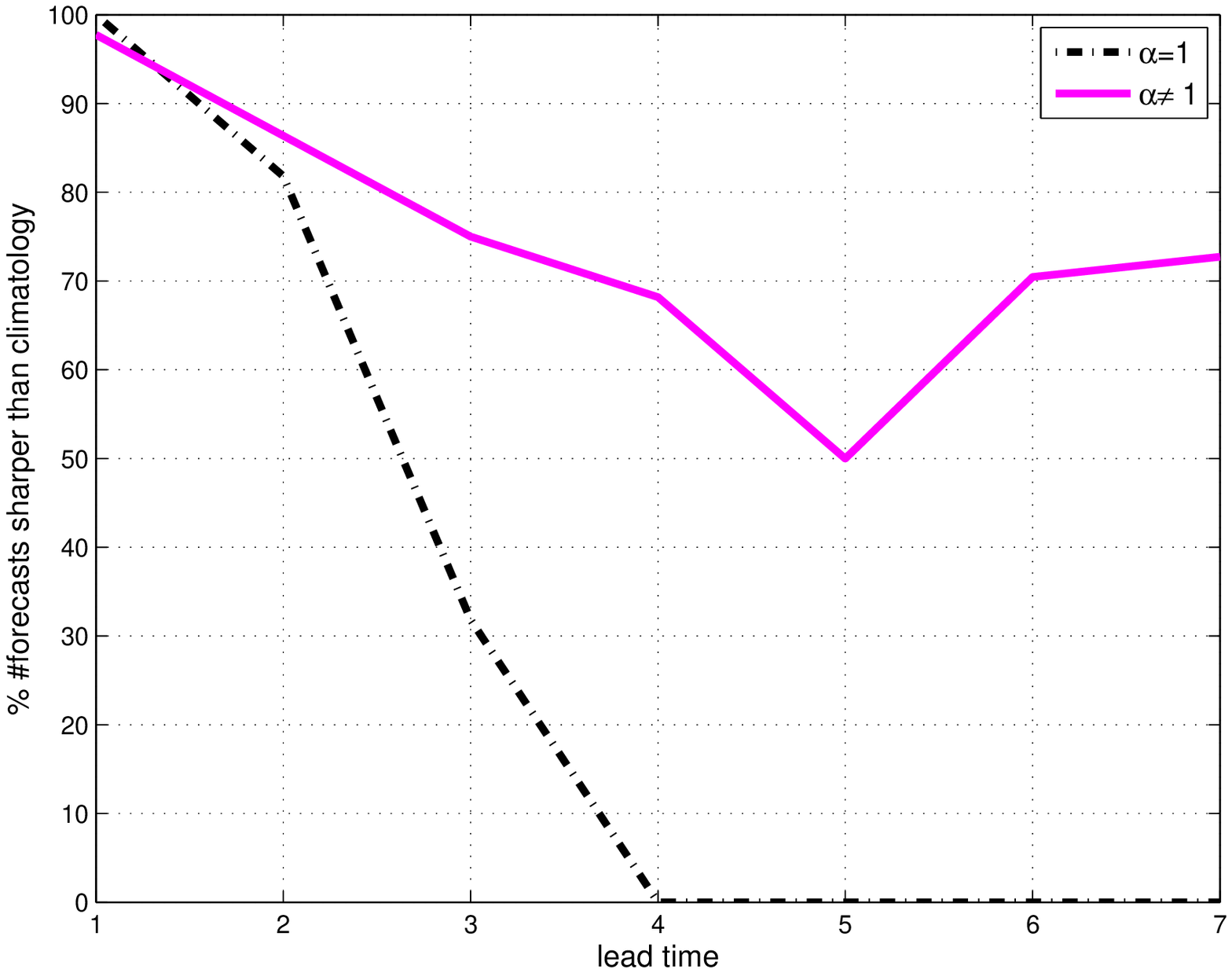,width=7.5cm,height=7.5cm},
\epsfig{file=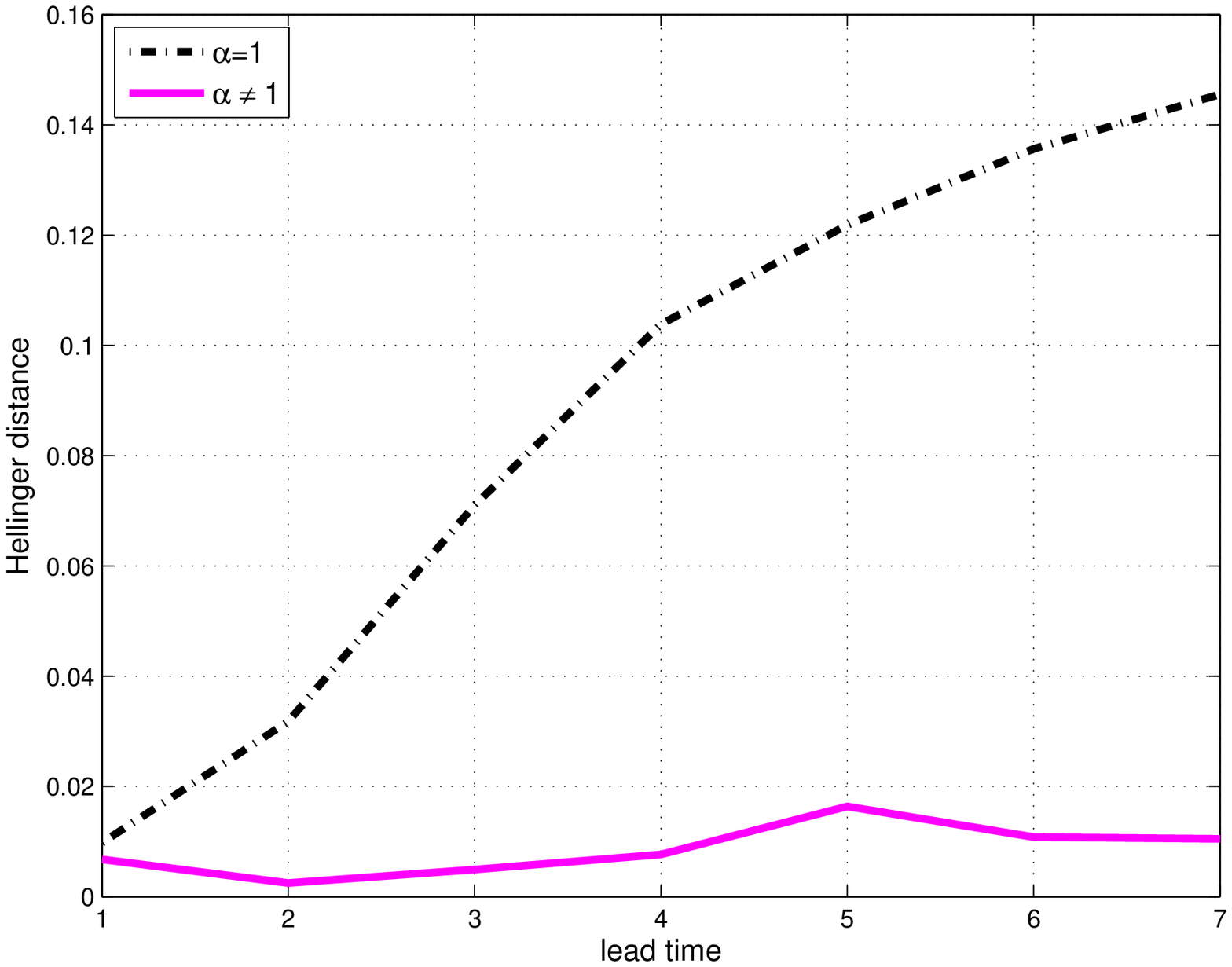,width=7.5cm,height=7.5cm}
}
\caption{\small\em (left) Graphs of percentage number of predictive distributions sharper than the unconditional distribution at various lead times and (right) graphs of Hellinger distance of forecaster's unconditional densities from the unconditional distribution versus lead time.}
\label{res:clim}
\end{figure}

A pair of density forecasts is shown in figure~\ref{res:rpdfs} to illustrate the effect on sharpness of mixing BOE forecasts with the unconditional density according to~(\ref{eqn:mix1}). Visual inspection suggests an increase in sharpness due to mixing with the unconditional density. Using entropy to measure sharpness, the forecasts were compared with the unconditional distribution. A graph of the percentage of forecasts sharper than the unconditional distribution against lead time is shown in figure~\ref{res:clim} on the left. From the graph it is evident that predictive distributions as issued by BOE are all less sharp than the unconditional distribution from as early as four quarters and ahead. This undermines the value of BOE fan charts from one year ahead and above in favour of the unconditional forecaster. At a lead time of one quarter ahead, it appears that mixing with the unconditional density results in mixture distributions that are less sharp than BOE forecasts. This, however, was at the expense of marginal calibration as is evident on the right hand graphs in figure~\ref{res:clim}.

\begin{figure}[tp]
\centering
\hbox{
\epsfig{file=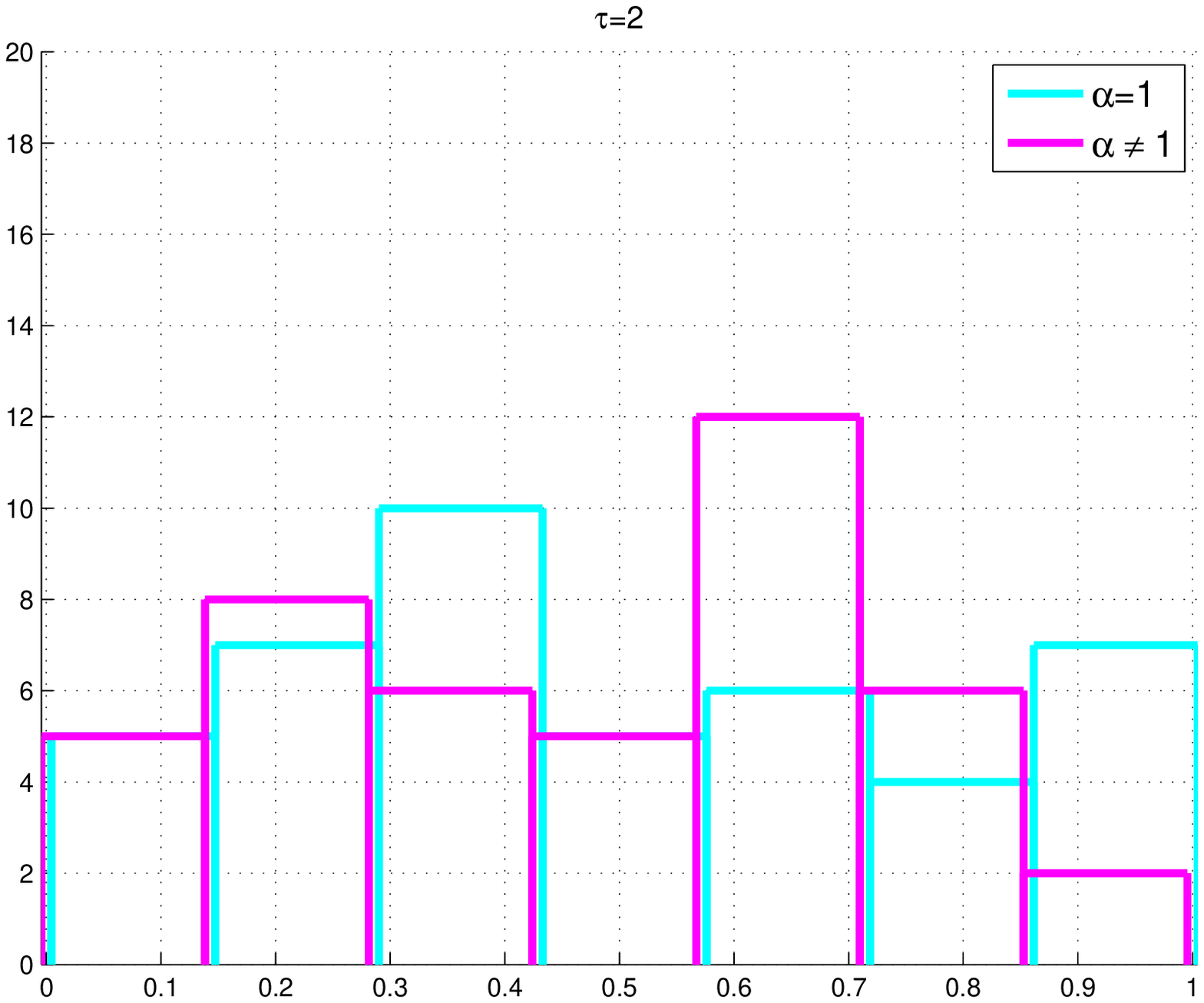,width=7.5cm,height=7.5cm},
\epsfig{file=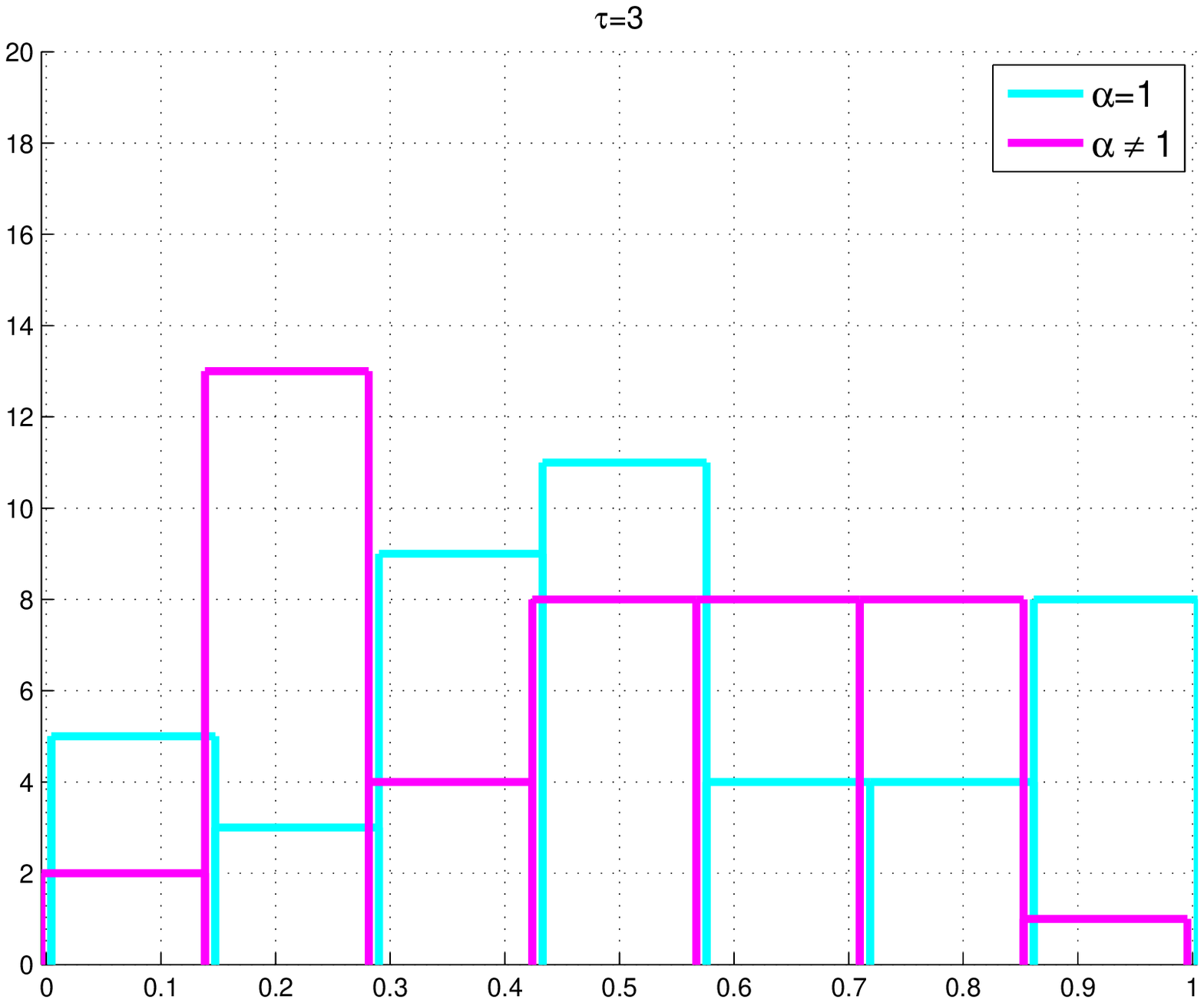,width=7.5cm,height=7.5cm}
}
\caption{\small\em (left) Distributions of PITs for RPIX inflation forecasts at lead times of 2 and 3 quarters ahead.}
\label{res:prob}
\end{figure}

The graphs of Hellinger distance of forecaster's unconditional densities from the unconditional densities in figure~\ref{res:clim} highlight the effect, on marginal calibration, of mixing with the unconditional density. At all the lead times considered, it is clear that mixing with the unconditional densities out-performed the BOE forecasts. Moreover, there is gain with respect to marginal calibration. Concerning probabilistic calibration, we consider only the lead times of two and three quarters ahead. The corresponding PIT graphs are shown in figure~\ref{res:prob}. Visual inspection suggests that mixing with the unconditional density does not have a significant effect on the quality of the PITs.
\section{Discussion and Conclusions}
\label{sec:con}
This paper presented a new theoretical and empirical analysis of the quality of density forecasts in terms of sharpness and calibration. It revisited the conjecture of \cite{til-gne} that a sufficiently calibrated forecaster is no more sharper than the ideal forecaster and proved a relevant proposition. It turned out that one cannot have both probabilistic and marginal calibration hold when the underlying model is mis-specified unless they settle for the unconditional distribution. Therefore, the paper argued for scaling down calibration expectations when facing model mis-specification. It focused upon combining conditional forecasts with the unconditional density. 

It was found that including the unconditional density via the logarithmic scoring rule tended to improve marginal calibration and maintain probabilistic calibration. This could be accompanied by a corresponding increase in sharpness as measured by entropy. Improvement in marginal calibration increased with lead time with no obvious compromise to probabilistic calibration. Fairly calibrated predictive distributions at higher lead times were found to be generally sharper than the unconditional distribution, thus affording early warning. Crucially, though, some of the density forecasts may have larger entropy than the unconditional distribution. Such forecasts may have to be rejected in favour of the unconditional distribution, which is sharper. These observations were made on RPIX inflation forecasts issued by the BOE. The forecasting model was nonlinear and mis-specified, having some stochastic component.

Relative to the unconditional distribution of RPIX inflation data, we found the BOE density forecasts to be very pessimistic, especially from lead times of four quarters and above. At lead times above three quarters, the unconditional distribution was found to be sharper than all the BOE forecasts. This undermines the value of the BOE forecasts at these lead times. Taking into account the BOE's judgement, mixing conditional forecasts with the unconditional density was found to yield sharper forecasts than the unconditional distribution. For instance, at a lead time of three quarters, only 30\% of the BOE density forecasts were sharper than the unconditional distribution. Mixing with the unconditional density improved sharpness of the forecasts so that about 75\% of them became sharper than the unconditional distribution.

It is useful to note that our calculations based on the BOE forecasts were performed on only 44 data points. This indicates that the methodology is relevant to data poor situations. If the system was not stationary, one would need to transform the time series to ensure stationarity. The methodology is applicable to both linear and nonlinear, stochastic systems provided stationarity has been established. Nonlinearity abounds in finance with examples in stock markets~\citep{kan-03,lind-93}. In macro-economics, GDP forecasting is an immediate example that would also be profited by this study.
\section*{Acknowledgements}
I would like to acknowledge useful discussions with members of Centre for the Analysis of Time Series at the London School of Economics, the Applied Dynamical Systems and Inverse Problems group at Oxford, Prof T. Gneiting, and Dr. D. J. Allwright. Comments from the Associate Editor and anonymous referees have helped greatly improve this manuscript. This work was supported by the RCUK Digital Economy Programme via EPSRC grant EP/G065802/1 The Horizon Digital Economy Hub.
\bibliographystyle{jofbib}
\bibliography{refs}

\begin{thebibliography}{55}
\providecommand{\natexlab}[1]{#1}
\providecommand{\url}[1]{\texttt{#1}}
\providecommand{\urlprefix}{URL }
\expandafter\ifx\csname urlstyle\endcsname\relax
  \providecommand{\doi}[1]{doi:\discretionary{}{}{}#1}\else
  \providecommand{\doi}{doi:\discretionary{}{}{}\begingroup
  \urlstyle{rm}\Url}\fi

\bibitem[{Bates and Granger(1969)}]{bat-69}
Bates JM, Granger CWJ. 1969.
\newblock The {C}ombination of {F}orecasts.
\newblock \emph{Operational Research Society} \textbf{20}:451--468.

\bibitem[{Benati(2004)}]{ben-04}
Benati L. 2004.
\newblock {Evolving Post-World War II U.K. Economic Performance}.
\newblock \emph{Journal of Money, Credit and Banking} \textbf{36}:691--717.

\bibitem[{Boero \emph{et~al.}(2008)Boero, Smith and Wallis}]{boe-08}
Boero G, Smith J, Wallis KF. 2008.
\newblock Modelling {UK} inflation uncertainty, 1958-2006.
\newblock In Watson M, Bollerslev T, Russell J, editors, \emph{Volatility and
  {T}ime {S}eries {E}conometrics: {Essays in Honour of Robert R. Engle}}.
  Oxford, UK: Oxford University Press.

\bibitem[{Brier(1950)}]{brie}
Brier GW. 1950.
\newblock Verification of forecasts expressed in terms of probability.
\newblock \emph{Monthly Weather Review} \textbf{78}:1--3.

\bibitem[{Britton \emph{et~al.}(1998)Britton, Fisher and Whitley}]{bri-98}
Britton E, Fisher P, Whitley J. 1998.
\newblock Inflation {R}eport projections: understanding the fan chart.
\newblock \emph{Bank of England Quarterly Bulletin} .

\bibitem[{Brocker(2009)}]{joc-09}
Brocker J. 2009.
\newblock Reliability, sufficiency, and the decomposition of proper scores.
\newblock \emph{Quarterly Journal of the Royal Meteorological Society}
  \textbf{135}:1512--1519.

\bibitem[{Brocker and Smith(2007)}]{joc-06}
Brocker J, Smith LA. 2007.
\newblock Scoring {P}robabilistic {F}orecasts: {T}he importance of being
  proper.
\newblock \emph{Weather and Forecasting} \textbf{22}:382--388.

\bibitem[{Brocker and Smith(2008)}]{joc-07}
Brocker J, Smith LA. 2008.
\newblock From ensemble forecasts to predictive distribution functions.
\newblock \emph{Tellus A} \textbf{60}:663.

\bibitem[{Bross(1953)}]{bross-53}
Bross IDJ. 1953.
\newblock \emph{Design for Decision: an introduction to statistical
  decision-making}.
\newblock New York: Macmillan.

\bibitem[{Bunn(1989)}]{bun-89}
Bunn D. 1989.
\newblock Forecasting with more than one model.
\newblock \emph{Journal of Forecasting} \textbf{8}:161--166.

\bibitem[{Casella(1985)}]{cas-85}
Casella G. 1985.
\newblock An {I}ntroduction to {E}mpirical {B}ayes {D}ata {A}nalysis.
\newblock \emph{The American Statistician} \textbf{39}:83--87.

\bibitem[{Clemen(1989)}]{cle-89}
Clemen RT. 1989.
\newblock Combining forecasts: {A} review and annotated bibliography.
\newblock \emph{International Journal of Forecasting} \textbf{5}:559--583.

\bibitem[{Clemen and Winkler(1999)}]{cle-99}
Clemen RT, Winkler RL. 1999.
\newblock Combining {P}robability {D}istributions {F}rom {E}xperts in {R}isk
  {A}nalysis.
\newblock \emph{Risk Analysis} \textbf{19}:187--203.

\bibitem[{Clements and Smith(2001)}]{clem-01}
Clements MP, Smith J. 2001.
\newblock Evaluating forecasts from {SETAR} models of exchange rates.
\newblock \emph{Journal of International Money and Finance}
  \textbf{20}:133--148.

\bibitem[{Corradi and Swanson(2006)}]{gra-06}
Corradi V, Swanson NR. 2006.
\newblock Predictive density evaluation.
\newblock In Elliott G, Granger CWJ, Timmermann A, editors, \emph{Handbook for
  {E}conometric {F}orecasting}, volume~1. North-Holland, 197--284.

\bibitem[{Dawid(1984)}]{daw-84}
Dawid AP. 1984.
\newblock Present position and potential developments: {S}ome {P}ersonal
  {V}iews: {S}tatistical {T}heory: {T}he {P}requencial {A}pproach.
\newblock \emph{J. R. Statist. Soc. A} \textbf{147}:278--292.

\bibitem[{Diebold \emph{et~al.}(1998)Diebold, Gunther and Tay}]{die-98}
Diebold FX, Gunther TA, Tay AS. 1998.
\newblock Evaluating density forecasts with application to financial risk
  management.
\newblock \emph{International Economic Review} \textbf{39}:863--883.

\bibitem[{Dowd(2007)}]{dow-07}
Dowd K. 2007.
\newblock Too good to be true? {T}he ({In})credibility of the {UK} inflation
  fan charts.
\newblock \emph{Journal of Macroeconomics} \textbf{29}:91--102.

\bibitem[{Efron and Morris(1977)}]{efr-77}
Efron B, Morris C. 1977.
\newblock Stein's {P}aradox in {S}tatistics.
\newblock \emph{Scientific American} \textbf{236}:119--127.

\bibitem[{Gneiting(2008)}]{gne-08}
Gneiting T. 2008.
\newblock Probabilistic forecasting.
\newblock \emph{J. R. Statist. Soc A} \textbf{171}:319--321.

\bibitem[{Gneiting(2011)}]{gne-11}
Gneiting T. 2011.
\newblock Quantiles as optimal point forecasts.
\newblock \emph{International Journal of Forecasting} \textbf{27}:197--207.

\bibitem[{Gneiting \emph{et~al.}(2007)Gneiting, Balabdaoui and
  Raftery}]{til-gne}
Gneiting T, Balabdaoui F, Raftery AE. 2007.
\newblock Probabilistic forecasts, calibration and sharpness.
\newblock \emph{J. R. Statist. Soc. B} \textbf{69}:243--268.

\bibitem[{Gneiting and Raftery(2007)}]{til-07}
Gneiting T, Raftery AE. 2007.
\newblock Strictly proper scoring rules, prediction and estimation.
\newblock \emph{J. Amer. Math. Soc.} \textbf{102}:359--378.

\bibitem[{Good(1952)}]{good-52}
Good IJ. 1952.
\newblock Rational decisions.
\newblock \emph{Journal of the Royal Statistical Society. Series B
  (Methodological)} \textbf{14}:107--114.

\bibitem[{Granger and Ramanathan(1984)}]{gra-ram}
Granger CWJ, Ramanathan R. 1984.
\newblock Improved methods of combining forecasts.
\newblock \emph{Journal of Forecasting} \textbf{3}:197--204.

\bibitem[{Greis and Gilstein(1991)}]{grei-91}
Greis NP, Gilstein CZ. 1991.
\newblock {Empirical Bayes methods for telecommunications forecasting}.
\newblock \emph{International Journal of Forecasting} \textbf{7}:183--197.

\bibitem[{Hall and Mitchell(2007)}]{hal-07}
Hall SG, Mitchell J. 2007.
\newblock Combining density forecasts.
\newblock \emph{International Journal of Forecasting} \textbf{23}:1--13.

\bibitem[{Harrison \emph{et~al.}(2005)Harrison, Nikolov, Quinn, Ramsay, Scott
  and Thomas}]{harr-05}
Harrison R, Nikolov K, Quinn M, Ramsay G, Scott A, Thomas R. 2005.
\newblock \emph{Bank of {E}ngland {Q}uarterly {M}odel}.
\newblock Bank of England.

\bibitem[{Hendry and Clements(2004)}]{hend-04}
Hendry DF, Clements MP. 2004.
\newblock Pooling forecasts.
\newblock \emph{Econometrics Journal} \textbf{7}:1--31.

\bibitem[{Hirschman(1957)}]{hir-57}
Hirschman II. 1957.
\newblock A note on entropy.
\newblock \emph{American Journal of Mathematics} \textbf{79}:152--156.

\bibitem[{James and Stein(1961)}]{jam-61}
James W, Stein C. 1961.
\newblock Estimation with {Q}uadratic {L}oss.
\newblock In \emph{Proc. Fourth Berkerly Symp. on Math. Statist. and Prob.}
  University of California Press, 361--379.

\bibitem[{Kanas(2003)}]{kan-03}
Kanas A. 2003.
\newblock Non-linear {F}orecasts of {S}tock {R}eturns.
\newblock \emph{Journal of Forecasting} \textbf{22}:299--315.

\bibitem[{Kapetanios \emph{et~al.}(2008)Kapetanios, Labhard and Price}]{kap-08}
Kapetanios G, Labhard V, Price S. 2008.
\newblock {Forecast combination and the Bank of Engalnd's suite of statistical
  forecasting models}.
\newblock \emph{Economic Modelling} \textbf{25}:772--792.

\bibitem[{Kelly(1956)}]{kel-56}
Kelly JL. 1956.
\newblock A new interpretation of information rate.
\newblock \emph{The Bell Systems Technical Journal} \textbf{35}:916--926.

\bibitem[{Knorr-Held and Rainer(2001)}]{knor-01}
Knorr-Held L, Rainer E. 2001.
\newblock Projections of lung cancer mortality in {W}est {G}ermany: A case
  study in {B}ayesian prediction.
\newblock \emph{Biostatistics} \textbf{2}:109--129.

\bibitem[{Kullback and Leibler(1951)}]{kul-lei}
Kullback S, Leibler RA. 1951.
\newblock On {I}nformation and {S}ufficiency.
\newblock \emph{The {A}nnals of {M}athematical {S}tatistics}
  \textbf{22}:79--86.

\bibitem[{Lawrence \emph{et~al.}(2006)Lawrence, Goodwin, O'Connor and
  Onkal}]{law-06}
Lawrence M, Goodwin P, O'Connor M, Onkal D. 2006.
\newblock Judgmental forecasting: A review of progress over the last 25 years.
\newblock \emph{International Journal of Forecasting} \textbf{22}:493--518.

\bibitem[{Leutbecher and Palmer(2008)}]{leu-08}
Leutbecher M, Palmer TN. 2008.
\newblock Ensemble {F}orecasting.
\newblock \emph{Journal of Computational Physics} \textbf{227}:3515--3539.

\bibitem[{Linden \emph{et~al.}(1993)Linden, Satchell and Yoon}]{lind-93}
Linden N, Satchell S, Yoon Y. 1993.
\newblock Predicting british financial indices: An approach based on chaos
  theory.
\newblock \emph{Structual Change and Economic Dynamics} \textbf{4}:145--162.

\bibitem[{Murphy(1993)}]{mur-93}
Murphy AH. 1993.
\newblock What is a good forecast? {A}n essay on the nature of goodness in
  weather forecasting.
\newblock \emph{Weather and Forecasting} \textbf{8}:281--293.

\bibitem[{Murphy and Wilks(1998)}]{mur-98}
Murphy AH, Wilks DS. 1998.
\newblock A case study of the use of statistical models in forecast
  verification: Precipitation probability forecasts.
\newblock \emph{Weather and Forecasting} \textbf{13}:795--810.

\bibitem[{Pal(2009)}]{pal-09}
Pal S. 2009.
\newblock A note on a conjectured sharpness principle for probabilistic
  forecasting with calibration.
\newblock \emph{Biometrika} \textbf{96}:1019--1023.

\bibitem[{Pal(2010)}]{pal-10}
Pal S. 2010.
\newblock Amendments and {C}orrections: On a conjectured sharpness principle
  for probabilistic forecasting with calibration.
\newblock \emph{Biometrika} \textbf{97}:1013.

\bibitem[{Parzen(1962)}]{par-62}
Parzen E. 1962.
\newblock On the {E}stimation of a {P}robability {D}ensity {F}unction and
  {M}ode.
\newblock \emph{The {A}nnals of {M}athematical {S}tatistics}
  \textbf{33}:1065--1076.

\bibitem[{Pollard(2002)}]{pol-02}
Pollard D. 2002.
\newblock \emph{A User's Guide to Measure Theoretic Probability}.
\newblock Cambridge University Press.

\bibitem[{Raftery \emph{et~al.}(2005)Raftery, Gneiting, Balabdaoui and
  Polakowski}]{raf-05}
Raftery AE, Gneiting T, Balabdaoui F, Polakowski M. 2005.
\newblock Using {B}ayesian {M}odel {A}veraging to {C}alibrate {F}orecast
  {E}nsembles.
\newblock \emph{Monthly Weather Review} \textbf{133}:1155--1174.

\bibitem[{Roulston and Smith(2002)}]{roul-02}
Roulston MS, Smith LA. 2002.
\newblock Evaluating {P}robabilistic {F}orecasts {U}sing {I}nformation
  {T}heory.
\newblock \emph{Monthly {W}eather {R}eview} \textbf{130}:1653--1660.

\bibitem[{Saunders(1958)}]{sau-58}
Saunders F. 1958.
\newblock The evaluation of subjective probability forecasts.
\newblock Cambridge, Massachusetts Institute of Technology, Department of
  Meteorology, Contact AF 19(604)-1305, Sci. Rept. 5.

\bibitem[{Shannon(1948)}]{sha}
Shannon CE. 1948.
\newblock A {M}athematical theory of communication.
\newblock \emph{The Bell Systems Technology Journal}
  \textbf{27}:379--423,623--656.

\bibitem[{Shannon(1949)}]{sha-49}
Shannon CE, editor. 1949.
\newblock \emph{Communication in the presence of noise}, volume~37. Pro.
  Institute of Radio Engineers.

\bibitem[{Silverman(1986)}]{sil-86}
Silverman BW. 1986.
\newblock \emph{Density {E}stimation for {S}tatistics and {D}ata {A}nalysis}.
\newblock Chapman and Hall, first edition.

\bibitem[{Wallis(2005)}]{wal-05}
Wallis KF. 2005.
\newblock Combining {D}ensity and {I}nterval {F}orecasts: {A} {M}odest
  {P}roposal.
\newblock \emph{Oxford Bulletin of Economics and Statistics, 67, Supplement
  (2005) 0305--9049} .

\bibitem[{White(1982)}]{whi-82}
White H. 1982.
\newblock Maximum likelihood estimation of misspecified models.
\newblock \emph{Econometrica} \textbf{50}:1--25.

\bibitem[{White(1994)}]{whi-94}
White H. 1994.
\newblock \emph{Estimation, Inference and Specification Analysis}.
\newblock New York: Cambridge University Press.

\bibitem[{Wilks(2006)}]{wilk-06}
Wilks DS. 2006.
\newblock \emph{Statistical Methods in the Atmospheric Sciences}.
\newblock Academic Press, 2nd ed edition.

\end{thebibliography}
\appendix
\numberwithin{equation}{section}
\section{Generalised Construction of Probabilistically Calibrated Forecasts }
\label{app:cal}
\begin{proposition}
\label{gencal}
Suppose that $G_t$ is a continuous strictly increasing distribution function on an interval $I_t$. Let $I$ be any interval and choose for each $t$ a strictly increasing continuous map $h_t:I\rightarrow I_t$. A probabilistically calibrated forecast distribution function precisely takes the form
\begin{equation}
F_s(x_s)=\frac{1}{T}\sum_{t=1}^TG_t\left[h_t\left\{h_s^{-1}(x_s)\right\}\right].
\label{gen:eqn}
\end{equation}
\end{proposition}
Equation~(\ref{gen:eqn}) is just \cite{til-gne}'s construction in \S~2.4 except that $T$ is general rather than 2 and the linear maps $x$ and $x/a$ that \cite{til-gne} use are replaced by the nonlinear maps $h_t$. Note that each $F_s$ is a strictly increasing continuous distribution on $I_s$ and they are probabilistically calibrated forecasts of the $G_t$'s, because given $0<p<1$ there is some $x$ in $I$ with
\begin{equation*}
p=\frac{1}{T}\sum_{t=1}^TG_t\left\{h_t(x)\right\}=F_s\left\{h_s(x)\right\},
\end{equation*}
whence $F_t^{-1}(p)=h_t(x)$ and
\begin{equation*}
\frac{1}{T}\sum_{t=1}^TG_t\left\{F_t^{-1}(p)\right\}=\frac{1}{T}\sum_{t=1}^TG_t\left\{h_t(x)\right\}=p.
\end{equation*}

Moreover, any probabilistically calibrated forecast of $G_t$ takes exactly this form. To see this, let $I$ be any interval and $h_1$ be any suitable map from $I$ onto $I_1$ and then define $h_t(x)=F_t^{-1}[F_1\{h_1(x)\}]$. It then follows that
\begin{equation*}
\frac{1}{T}\sum_{t=1}^TG_t\left[h_t\left\{h_s^{-1}(x_s)\right\}\right]=\frac{1}{T}\sum_{t=1}^TG_t\left[F_t^{-1}\left\{F_s(x_s)\right\}\right]=F_s(x_s).
\end{equation*}
The first equality follows by definition of the $h_t$ functions and the next by the probabilistic calibration property. Hence the $F_t$'s have exactly the form of the construction.
\section{Proof of Proposition \ref{sharpness}}
\label{app:sha}
Using proposition~\ref{gencal}, probabilistic calibration implies that $F_t$ takes precisely the form given in (\ref{gen:eqn}). If the sequence $\{F_t\}$ is also finite marginally calibrated, we can substitute~(\ref{gen:eqn}) into~(\ref{eqn:mar}) to obtain 
\begin{eqnarray*}
\frac{1}{T(T-1)}\sum_{t=1}^T\sum_{s=1,s\neq t}^TG_t\left[h_t\left\{h_s^{-1}(x)\right\}\right]=\frac{1}{T}\sum_{t=1}^TG_t(x),\quad T\ge2.
\end{eqnarray*}
It is, therefore, required that $G_t\left[h_t\left\{h_s^{-1}(x)\right\}\right]=G_{i}(x)$, where $i\in\{1,..,T\}$. If $G_t[h_t\{h_s^{-1}(x)\}]=G_s(x)$ for any $s$, then the forecasts $\{F_t\}$ are ideal. On the other hand, if $G_t[h_t\{h_s^{-1}(x)\}]=G_t(x)$, then we have the finite unconditional forecaster (FUF). 

We now wish to show that a non-FUF forecaster who is both probabilistically and marginally calibrated is precisely the ideal forecaster. Consider $F_s(x)$ as defined by equation~(\ref{gen:eqn}) for a given $s$. Suppose there exists $q$ such that 
\begin{equation} 
G_t[h_t\{h_s^{-1}(x)\}]=G_s(x), \quad\mbox{for all}\quad t\le q
\label{app:gs}
\end{equation} 
and 
\begin{equation}
G_t[h_t\{h_s^{-1}(x)]=G_t(x) \quad\mbox{for all}\quad t>q.
\label{app:eql}
\end{equation}
Equation~(\ref{app:gs}) implies that $G_s[h_s\{h_t^{-1}(x)\}]=G_t(x)$ for all $t\le q$ while (\ref{app:eql}) implies that $h_t(x)=h_s(x)$ for all $t>q$. $F_s(x)$ contains $q$ counts of $G_s(x)$.  Each
\begin{equation*}
F_i(x)=\frac{1}{T}\sum_{t=1}^TG_t\left[h_t\left\{h_i^{-1}(x)\right\}\right],
\end{equation*}
$i\neq s$, contains 0 counts of $G_s(x)$ if $i\le q$. If $i>q$, we get 
\begin{equation*}
G_t[h_t\{h_i^{-1}(x)\}]=G_t[h_t\{h_s^{-1}(x)\}]=G_s(x),
\end{equation*}
for all $t\le q$. The first equality follows from noting that $h_i(x)=h_s(x)$ and the second from applying~(\ref{app:gs}). Hence each $F_i(x)$ contains $q$ counts of $G_s(x)$. Therefore, all the summations on the right hand side of the forecasters contain $q+(T-q)q$ counts of $G_s(x)$. Finite marginal calibration imposes the requirement that $q+(T-q)q=T$, which holds if and only if $q=T$. But $q=T$ implies that we have ideal forecasts.

More generally, the sequence $\{G_t[h_t\{h_s^{-1}(x)\}]\}_{t>q}$ may contain multiplicities of the $G_t(x)$ terms. This means that, for a given $t=r>q$ for which $G_r[h_r\{h_s^{-1}(x)\}]=G_r(x)$, there may be at least another $p\neq r$ and $p>q$ such that $G_p[h_p\{h_s^{-1}(x)\}]=G_r(x)$. Let $j$ be the number of all $p$'s for all $r$'s as defined above. Then the total number of $G_s(x)$ terms over the right hand sides of all $F_i(x)$ and $F_s(x)$ is $q+(T-q-j)q$. Marginal calibration imposes the condition that
\begin{equation*}
q+(T-q-j)q=T\quad\Rightarrow\quad q^2-q(T-j+1)+T=0.
\end{equation*}
For the above quadratic equation to have an integer solution in $q$, the discriminant must be a perfect square, which happens if and only if $j=0$. Hence a non-FUF forecaster who is both finite marginally and probabilistically calibrated must have issued ideal forecasts. 
\section{Proofs of Propositions~\ref{prop:returns} and~\ref{prop:discount}}
\label{app:proofs}
{\bf Proof of proposition~\ref{prop:returns}:} The second partial derivative of equation~(\ref{eqn:ei5}) with respect to the mixture parameter $\alpha$ yields
\begin{equation*}
\frac{\partial^2\langle\mbox{IGN}\rangle}{\partial\alpha^2}=\frac{1}{T}\sum_{t=1}^T\left\{\frac{\rho^{(t)}(s_t|\mathcal{V}_T)-\rho_c(s_t)}{f^{(t)}(s_t|\mathcal{V}_T)}\right\}^2.
\end{equation*} 
Hence the first derivative of $\langle\mbox{IGN}\rangle$ with respect to $\alpha$ is an increasing function of $\alpha$. It follows that the first derivative will have a zero at some $\alpha=\alpha_*\in(0,1)$ if and only if
\begin{equation*}
\left.\frac{\partial\langle\mbox{IGN}\rangle}{\partial\alpha}\right|_{\alpha=0}<0\quad\mbox{and}\quad\left.\frac{\partial\langle\mbox{IGN}\rangle}{\partial\alpha}\right|_{\alpha=1}>0.
\end{equation*}
These are essentially the inequalities in the proposition. The second derivative implies that $\alpha_*$ is a global minimiser of the score.

{\bf Proof of proposition~\ref{prop:discount}:} $\partial\langle\mbox{IGN}\rangle/\partial\sigma=0$ implies that 
\begin{equation*}
\sigma_*^2\frac{1}{T}\sum_{t=1}^T\frac{\rho^{(t)}(s_t|\mathcal{V}_T)}{f^{(t)}(s_t|\mathcal{V}_T)}=\frac{1}{T}\sum_{t=1}^T\left\{s_t-X_1^{(t)}\right\}^2\frac{\rho^{(t)}(s_t|\mathcal{V}_T)}{f^{(t)}(s_t|\mathcal{V}_T)}.
\end{equation*}
But $\partial\langle\mbox{IGN}\rangle/\partial\alpha=0$ implies that
\begin{equation*}
\frac{1}{T}\sum_{t=1}^T\frac{\rho^{(t)}(s_t|\mathcal{V}_T)}{f^{(t)}(s_t|\mathcal{V}_T)}=1,
\end{equation*}
which may be plugged into the left hand side of the previous equation to complete the proof.
\section{Combining point forecasts}
\label{app:point}
Consider two forecasting models each with standard deviation of errors given by $\sigma_1$ and $\sigma_2$, respectively. Further more, suppose the correlation coefficient of the forecasting errors of these models is  $\rho$. We assume that forecasting errors of each model are not biased, otherwise the modeller can always correct the bias. If we make a forecast combination, $y_c=\alpha y_1+(1-\alpha)y_2$, of models $1$ and $2$, then the following proposition, which is a counter part of Proposition~\ref{prop:returns}, holds:
\begin{proposition}
\label{prop:point}
Suppose that the standard deviations of forecasting errors of two models are $\sigma_1,\sigma_2\neq 0$, respectively. If the forecast errors are not biased, then the necessary and sufficient condition for improvement of the combined forecast in the sense of mean squared errors is
\begin{equation}
\rho<\min\left\{\frac{\sigma_2}{\sigma_1},\frac{\sigma_1}{\sigma_2}\right\},
\label{eqn:appd1}
\end{equation}
where $\rho$ is the correlation coefficient of the forecasting errors.
\end{proposition}
Proof: Suppose the forecast errors of each model are $e_1$ and $e_2$, respectively. Then the forecast error of the combined model is $e_c=\alpha e_1+(1-\alpha)e_2$. Since we assumed the errors are not biased, it follows that $\mathbb{E}[e_i]=0$, $i=1,2$. Therefore, $\sigma_i^2=\mathbb{E}[e_i^2]$, $i=1,2$. The mean squared errors of the combined forecast then satisfy the relation $\sigma_c^2=\mathbb{E}[e_c^2]$, whence
\begin{equation}
\sigma_c^2=\alpha^2\sigma_1^2+(1-\alpha)^2\sigma_2^2+2\rho\alpha(1-\alpha)\sigma_1\sigma_2.
\label{eqn:appd3}
\end{equation}
Differentiating~(\ref{eqn:appd3}) with respect to $\alpha$ yields
\begin{equation}
\frac{\ud\sigma_c^2}{\ud\alpha}=2\alpha(\sigma_1^2+\sigma_2^2-2\rho\sigma_1\sigma_2)-2(\sigma_2^2-\rho\sigma_1\sigma_2).
\label{eqn:appd4}
\end{equation}
At the extremum of the variance $\sigma_c^2$, $\ud\sigma_2^2/\ud\alpha=0$, which yields
\begin{equation}
\alpha^*=\frac{\sigma_2^2-\rho\sigma_1\sigma_2}{\sigma_1^2+\sigma_2^2-2\rho\sigma_1\sigma_2}.
\label{eqn:appd5}
\end{equation}
In order to deduce that $\alpha^*$ is a global minimum, it suffices to note that
\begin{equation*}
\frac{\ud^2\sigma_c^2}{\ud\alpha^2}=2(\sigma_1^2+\sigma_2^2-2\rho\sigma_1\sigma_2)>0.
\end{equation*}
Condition~(\ref{eqn:appd1}) follows upon imposing the requirement that $0<\alpha*<1$.

Note that if either variance of the forecast error vanishes, there is no need for model combination. The above proposition together with its proof lead us to the following corollary:
 \newtheorem{corollary}{Corollary}
\begin{corollary}
\label{app:cor}
If condition~(\ref{eqn:appd1}) holds, then the variance of the forecast errors of the combined model, $\sigma_c^2$, $\alpha^*$ is smaller than that of either constituent model, i.e.
\begin{equation}
\sigma_c^2<\min\left\{\sigma_1^2,\sigma_2^2\right\}.
\label{eqn:appd2}
\end{equation}
\end{corollary}
The above corollary may be viewed as a point forecast parallel of Proposition~\ref{prop:discount}. In some sense it guarantees the `sharpness' of the combined forecast. Unfortunately, here sharpness has to be thought of as a property of both forecasts and observations.
\end{document}